\documentclass[12pt]{article}

\usepackage{a4}
\usepackage{epsfig}

\newlength{\abstwidth}
\def\be{\begin{equation}}
\def\ee{\end{equation}}

\begin{document}

\def\lsim{\mathrel{\rlap{\lower4pt\hbox{\hskip1pt$\sim$}}
    \raise1pt\hbox{$<$}}}         
\def\gsim{\mathrel{\rlap{\lower4pt\hbox{\hskip1pt$\sim$}}
    \raise1pt\hbox{$>$}}}         

\pagestyle{empty}

\begin{flushright}
BI-TH 2001/13\\
MPI-PhT/2001-23, July 16, 2001
\end{flushright}

\vspace{\fill}

\begin{center}
{\Large\bf Scaling in \boldmath{$\gamma^* \lowercase{p}$} total cross
sections, \\[1ex]
saturation and the gluon density ${}^{*}$}\\[1.8ex]
{\bf Dieter Schildknecht} \\[1.2mm]
Fakult\"{a}t f\"{u}r Physik, Universit\"{a}t Bielefeld \\[1.2mm]
D-33501 Bielefeld, Germany${}^{**}$ \\[1.2mm]
and MPI f\"ur Physik, M\"unchen, Germany\\[1.5ex]
{\bf Bernd Surrow} \\[1.2mm]
DESY, 22607 Hamburg, Germany \\[1.2mm]
and \\[1.5ex]
{\bf Mikhail Tentyukov${}^{***}$} \\[1.2mm]
Fakult\"{a}t f\"{u}r Physik, Universit\"{a}t Bielefeld \\[1.2mm]
D-33501 Bielefeld, Germany

\end{center}

\vspace{\fill}

\begin{center}
{\bf Abstract}\\[2ex]
\begin{minipage}{\abstwidth}
Including the new HERA data, the $\gamma^* p$ total cross section is
analysed in the generalized vector dominance/colour-dipole picture (GVD/CDP)
that contains scaling in $\eta = (Q^2 + m^2_0 ) / \Lambda^2 (W^2)$,
where $\Lambda^2 (W^2)$ is an increasing function of $W^2$.
 At any
$Q^2$, for $W^2 \rightarrow \infty$, the cross sections for virtual and real
photons become identical, $\sigma_{\gamma^* p} (W^2 , Q^2) / \sigma_{\gamma p}
(W^2) \rightarrow 1$. The gluon density deduced from the
colour-dipole cross section fulfills the leading order DGLAP relationship.
Evolution \`a la DGLAP breaks down for $\eta \lsim 0.1$.
\end{minipage}
\end{center}

\vspace{\fill}
\noindent

\vspace{\fill}
\noindent

\rule{60mm}{0.4mm}

\vspace{0.1mm}
\noindent
${}^*$ Supported by the BMBF, Contract 05 HT9PBA2\\
${}^{**}$ Permanent address\\
${}^{***}$ On leave from BLTP JINR, Dubna, Russia
\clearpage
\pagestyle{plain}
\setcounter{page}{1}

\baselineskip 20pt

Recently we have shown that the data for the total virtual-photon-proton
cross section in the low-x diffraction region $(x \simeq \frac{Q^2}{W^2} \ll
1)$ fulfill a scaling law. We found \cite{1}
\be
\sigma_{\gamma^* p} (W^2 , Q^2 ) = \sigma_{\gamma^* p} (\eta (W^2 , Q^2)) ,
\label{(1)}
\ee
where the dimensionless scaling variable $\eta$ is given by
\be
\eta (W^2 , Q^2 ) = \frac{Q^2+m^2_0}{\Lambda^2(W^2)},
\label{(2)}
\ee
and $\Lambda^2 ( W^2)$ is a (slowly) increasing function of $W^2$,
conveniently parameterized by the power law
\be
\Lambda^2 (W^2) = C_1 ( W^2 + W^2_0)^{C_2}
\label{(3)}
\ee
or by a logarithmic function of $W^2$.
The scaling behaviour (\ref{(1)})
was determined in a model-independent analysis of the
experimental data and subsequently it was interpreted \cite{1}
in terms of an ansatz \cite{3}
for the virtual-photon-proton cross section that was based
on the generic structure of two-gluon exchange \cite{2,4}
(compare Fig.1), the generalized vector dominance/colour-dipole picture
(GVD/CDP).
In the model-independent analysis of the data no specific ansatz for
the functional dependence of the cross section on $\eta$ (apart from
continuity in $\eta$) was introduced. The threshold mass, $m_0 < m_{\rho^0}$,
as well as the parameters $W^2_0$ and $C_2$\footnote{The absolute value of
  $C_1 > 0$ is irrelevant for the existence of the scaling behaviour. It only
  corresponds to a shift of the scale.}, were determined in a
least-squares fit to the experimental data. The theoretical ansatz for
$\sigma_{\gamma^* p}$, or rather the virtual-photon-proton
Compton-forward-scattering-amplitude in terms of (the generic structure of)
two-gluon exchange, the GVD/CDP, was shown to incorporate the observed
scaling, and a
comparison with the experimental data determined the fourth parameter
$C_2$.

The purpose of the present work is threefold,
\begin{itemize}
\item[i)]
to elaborate on the consequences of the observed scaling in $\eta$ with respect
to the asymptotic behaviour of $\sigma_{\gamma^* p}$
in the so-called ``saturation''
regime, $Q^2$ fixed, $W^2 \rightarrow \infty$, i.e.
$x\rightarrow 0$,
\item[ii)]
to include the more recently published data from ZEUS and H1 in the
analysis, and
\item[iii)]
to establish the connection between the above-mentioned description based
on the notion of the colour-dipole cross section and the conventional
gluon density of the proton.
\end{itemize}

The evaluation of the two-gluon exchange diagrams from Fig.1 in the
low-$x$ limit, and transition to transverse position space, leads to
\cite{2,1}
\begin{eqnarray}
\sigma_{\gamma^*p} (W^2, Q^2) & = &
\int dz \int d^2 r_\perp \vert \psi \vert^2
(r^2_\perp Q^2 z (1-z), Q^2 z (1-z), z) \cdot \nonumber \\
& \cdot &\sigma_{(q \bar q)p} (r^2_\perp, z(1-z),W^2).
\label{(4)}
\end{eqnarray}
where the Fourier representation of the colour-dipole cross section is
given by
\be
\sigma_{(q \bar q)p} (r^2_\perp, z (1-z),W^2)= \int d^2 l_\perp
\tilde \sigma_{(q \bar q)p} (\vec l^{~2}_\perp, z (1-z),W^2) \cdot
(1 - e^{- i \vec l_\perp \cdot \vec r_\perp}).
\label{(5)}
\ee
Indeed, substituting (\ref{(5)}) into (\ref{(4)}), together with the
Fourier representation of the photon-wave function $|\psi |^2$, yields
\cite{3}
precisely the momentum-space representation for the cross section based on
the diagrams in Fig.1 evaluated in the $x \rightarrow 0$ limit.
We note that the detailed structure of the interaction of the $(q \bar q)$
pair, once the pair has emitted (at least) two gluons, is fully buried in
the colour-dipole cross section, or rather in its Fourier-transform,
$\tilde\sigma_{(q \bar q)p}$, in (\ref{(5)}).

From (\ref{(5)}), for vanishing and for infinite quark-antiquark separation,
respectively, we have
\be
\sigma_{(q \bar q)p} (r^2_\perp, z (1-z),W^2) = \sigma^{(\infty)}
\cdot \left\{ \begin{array}{l@{\quad,\quad}l}
\frac{1}{4} r^2_\perp \langle \vec l^{~2} \rangle_{W^2,z}
& {\rm for}~ r^2_\perp
\to 0,\\
1 & {\rm for}~ r^2_\perp \to \infty,
\end{array}
\right.
\label{(6)}
\ee
where by definition
\be
\sigma^{(\infty)} = \pi \int dl^2_\perp \tilde \sigma (\vec l^2_\perp,
z (1-z),W^2),
\label{(7)}
\ee
and $\langle \vec l^2_\perp \rangle$, the average gluon transverse
momentum, is given by \cite{5}
\be
\langle \vec l^{~2} \rangle_{W^2,z} = \frac{\int d\vec l^{~2}_\perp
\vec l^{~2}_\perp
\tilde \sigma_{(q \bar q)p} (\vec l^{~2}_\perp, z(1-z),W^2)}
{\int d \vec l^{~2}_\perp
\tilde \sigma_{(q \bar q)p} (\vec l^{~2}_\perp, z(1-z),W^2)}.
\label{(8)}
\ee
If the GVD/CDP
defined by (\ref{(4)}) and (\ref{(5)}) is to be consistent, the
asymptotic $(\vec r^2_\perp \rightarrow \infty)$ limit of the
colour-dipole cross section, $\sigma^{(\infty)}$, defined in (\ref{(7)}),
must exist, and moreover it should be independent of the configuration
variable $z$.
In addition, $\sigma^{(\infty)}$ should at most be weakly dependent on $W^2$,
to fulfill the unitarity restrictions for hadron-hadron interactions for
$W^2\rightarrow\infty$.

In order to incorporate the requirement of scaling, (\ref{(1)}), we replace
$\vec r^2_\perp$ in (\ref{(4)}) by the dimensionless
variable \cite{5}
\be
u \equiv r^2_\perp \Lambda^2 (W^2) z (1-z).
\label{(9)}
\ee
The photon wave function in (\ref{(4)}) then becomes the function
$|\psi |^2 (u\frac{Q^2}{\Lambda^2}, \frac{Q^2}{\Lambda^2} , z)$.
Scaling in $\eta$ implies that the colour-dipole cross section solely
depends on $u$,
\be
\sigma_{(q \bar q)p} (r^2_\perp, z (1-z), W^2) = \sigma_{(q \bar q)p} (u),
\label{(10)}
\ee
and according to (\ref{(6)}), we have
\be
\langle \vec l^{~2}\rangle_{W^2,z} = \Lambda^2 (W^2) z (1-z).
\label{(11)}
\ee
Averaging over the configuration variable $z$ yields
\be
\langle \vec l^{~2} \rangle_{W^2} = \frac{1}{6} \Lambda^2 (W^2).
\label{(12)}
\ee
The quantity $\Lambda^2 (W^2)$ appearing in the scaling variable (\ref{(2)})
is accordingly (apart from the factor 1/6) identified as the average
transverse momentum of the gluon absorbed by the quark (or the antiquark).

It is illuminating to explicitly insert the average gluon transverse
momentum (\ref{(11)}) into (\ref{(6)}), yielding
\be
\sigma_{q \bar q p} = \sigma^{(\infty)} \cdot
\left\{ \begin{array}{l@{\quad,\quad}l}
\frac{1}{4} r^2_\perp \Lambda^2 (W^2) z (1-z) & {\rm for}~\Lambda^2 \cdot
r^2_\perp \to 0,\\
1 & {\rm for}~\Lambda^2 \cdot r^2_\perp \to \infty.
\end{array} \right.
\label{(13)}
\ee
Noting that the dependence of the photon wave function on $\vec r^2_\perp \cdot
Q^2$ requires small $\vec r^2_\perp$ at large $Q^2$ in order to develop
appreciable strength, an important qualitative conclusion on the
functional dependence on $\eta$ of $\sigma_{\gamma^* p} (\eta)$ can
immediately be drawn. For fixed energy and sufficiently large $Q^2$, the
dominant contribution to $\sigma_{\gamma^* p} (\eta)$ stems from the
$\Lambda^2 \cdot \vec r^2_\perp \rightarrow 0$ behaviour of the dipole
cross section in (\ref{(13)}). Increasing the energy $W^2$ and, accordingly,
$\Lambda^2 (W^2)$ at fixed $Q^2$, the $\Lambda^2 \cdot r^2_\perp \rightarrow
\infty$ behaviour of the dipole cross section in (\ref{(13)}) will
eventually become relevant, thus implying the transition from a fairly strong
$W^2$ dependence (due to $\Lambda^2 (W^2)$) to an increasingly weaker one.
This behaviour is precisely borne out by the data: The strong
dependence on $W^2$ at fixed $Q^2$ indeed turns into the weaker one of
photoproduction, compare Fig.2.
Thus, having supplemented the GVD/CDP, (\ref{(4)}),(\ref{(5)}),
by scaling in $\eta$, we qualitatively obtain the functional
behaviour of the experimental data in Fig.2.

For the quantitative evaluation of (\ref{(4)}) with (\ref{(5)}), we
approximated the (unknown) precise distribution in the gluon momentum
transfer $\vec l^2_\perp$ by a $\delta$-function situated at the
average gluon momentum (\ref{(11)}), i.e. \cite{1}
\be
\tilde \sigma_{(q \bar q)p} = \sigma^{(\infty)} \frac{1}{\pi} \delta
(\vec l^{~2}_\perp - \Lambda^2 (W^2) z (1-z)).
\label{(14)}
\ee
With (\ref{(14)}), and upon returning to momentum space, a complete analytical
evaluation of (\ref{(4)}) with (\ref{(5)}) was given. Ignoring corrections of
the order of $m^2_0/\Lambda^2 (W^2) \ll 1$, it reads \cite{1}\footnote{Compare
\cite{1} for the complete expression from the function $I_0 (\eta)$}
\be
\sigma_{\gamma^* p} (\eta) = \frac{\alpha R_{e^+ e^-}}{3\pi} \sigma^{(\infty)}
I_0 (\eta) \\
\cong \frac{2\alpha}{3\pi} \sigma^{(\infty)}
\left\{ \begin{array}{l@{\quad,\quad}l}
\ln (1/\eta) ,  & {\rm for}~ \eta \rightarrow 0, \\
1/2 \eta,  & {\rm for}~\eta \gg 1.
\end{array} \right.
\label{(15)}
\ee
For any arbitrary fixed value of $Q^2$, with $W^2 \rightarrow \infty$, the
soft logarithmic dependence of photoproduction and hadronic reactions as
a function of $\eta^{-1} = \Lambda^2 (W^2) / (Q^2 + m^2_0)$ is indeed reached.

From the logarithmic dependence on $\eta$ for $\eta\rightarrow 0$ in
(\ref{(15)}), we draw the important conclusion that virtual and real photons
become equivalent in the limit of $W^2 \rightarrow \infty$ (i.e. $x\rightarrow
0, Q^2$ fixed) \cite{5}
\be
\lim_{{W^2 \to \infty} \atop {Q^2 {\rm fixed}}} \frac{\sigma_{\gamma^*p}
(W^2,Q^2)}{\sigma_{\gamma p} (W^2)} = 1.
\label{(16)}
\ee
Note that the alternative of $\Lambda^2 = $ const. that implies Bjorken
scaling of the structure function $F_2 \sim Q^2 \sigma_{\gamma^* p}$ for
sufficiently large $Q^2$, leads to a result entirely different from
(\ref{(16)}),
\be
\lim_{{W^2 \to \infty} \atop {Q^2 {\rm fixed}}} \frac{\sigma_{\gamma^*p}
(W^2,Q^2)}{\sigma_{\gamma p} (W^2)} = \frac{\Lambda^2}{2 Q^2 \ln
\frac{\Lambda^2}{m^2_0}},~~({\rm assuming}~ \Lambda = const.),
\label{(17)}
\ee
i.e. a suppression of the virtual-photon-proton cross section
relative to photoproduction by a full power of $Q^2$ even in the asymptotic
limit of $W^2 \rightarrow \infty , Q^2$ fixed $(x \rightarrow 0)$.

In our reanalysis of the experimental data \cite{6,7},
we have replaced the H1 94 and the ZEUS 94 results by the most recent
H1 96/97 and ZEUS 96/97 data. The result for the fitted parameters differs
only within errors from the previous fit. This can be seen by comparing
our new results,
\begin{eqnarray}\label{(18)}
m^2_0 & = & 0.15 \pm 0.04~{\rm GeV}^2 \nonumber \\
C_1 & = & 0.34 \pm 0.06 , \nonumber \\
C_2 & = & 0.27 \pm 0.01 , \\
W^2_0 & = & 1081 \pm 124~{\rm GeV}^2 , \nonumber
\end{eqnarray}
with the ones in \cite{1}.
Figure 2 shows the data and the results from the GVD/CDP based on the
parameter set (\ref{(18)}).

We return to the above discussion on the asymptotic behaviour of
$\sigma_{\gamma^* p}$ given by (\ref{(16)}). The approach to this limit of
$x \rightarrow 0$ is governed by the limit of $\eta \rightarrow 0$ in Fig.2.
It is suggestive to consider the logarithmic derivative of $\ln \sigma_
{\gamma^* p}(\eta)$. From (\ref{(15)}), we have
\be
- \frac{d}{d \ln \eta} \ln \sigma_{\gamma^* p} (\eta) =
\frac{1}{I_0(\eta)} \eta \frac{d I_0 (\eta)}{d \eta} =
\left\{ \begin{array}{l@{\quad,\quad}l}
\frac{1}{\ln \eta^{-1}} ,  & {\rm for}~ \eta \rightarrow 0, \\
1 ,  & {\rm for}~\eta \rightarrow \infty.
\end{array} \right.
\label{(19)}
\ee
In Fig.3, we show the logarithmic derivative (\ref{(19)}) in conjuction with
the dependence of $W^2$ on $Q^2$ at fixed $\eta$. The ``saturation limit''
of a soft $W^2$ dependence ($Q^2$ arbitrary, fixed) requires $\eta \le 0.1$
The corresponding energy is extremely high, as may be read off from Fig.3b.

The approach of $\sigma_{\gamma^* p}$ to the limit (\ref{(16)}) is also
explicitly displayed in Fig.4 that shows $\sigma_{\gamma^* p}$ as a function
of $W^2$ for various fixed values of $Q^2$. We stress that the
limiting behaviour (\ref{(16)}), i.e. the universal cross section for virtual
and real photons, is the outgrowth of the $\ln (1/\eta)$ behaviour in
(\ref{(15)}) for $\eta \rightarrow 0$. This logarithmic functional
dependence is a
strict consequence of the GVD/CDP with scaling in $\eta$ that rests on the
underlying generic two-gluon-exchange structure from QCD in Fig.1.

So far we have exclusively concentrated on a representation of
$\sigma_{\gamma^* p}$ in terms of the colour-dipole cross section,
$\sigma_{(q \bar q) p} (\vec r^2_\perp , W^2 , z(1-z))$. For sufficiently large
$Q^2$ and non-asymptotic $W^{2}$, such that the $\Lambda^2 (W^2) \cdot \vec r^2_\perp \rightarrow 0$
limit in (\ref{(13)}) is valid, one may alternatively parameterize the lower
vertex in Fig.1 in terms of the gluon density of the proton. The corresponding
formula has indeed been worked out in \cite{8}. It reads
\be
\sigma_{(q \bar q) p} (r^2_\perp , x , Q^2) = \frac{\pi^2}{3} r^2_\perp x g (x,
Q^2 ) \alpha_s (Q^2) .
\label{(20)}
\ee
Identifying (\ref{(20)}) with the $\Lambda^2 (W^2) \cdot r^2_\perp \rightarrow 0$
form of $\sigma_{(q \bar q) p}$ from (\ref{(13)}), upon averaging over
$z (1-z)$ as in (\ref{(12)}),
\be
\bar\sigma_{(q \bar q) p} ( r^2_\perp , W^2) = \sigma^{(\infty)} \frac{1}{24}
r^2_\perp \Lambda^2 (W^2) ,
\label{(21)}
\ee
we deduce
\be
xg (x , Q^2) \alpha_s (Q^2) = \frac{1}{8\pi^2} \sigma^{(\infty)} \Lambda^2
\left( \frac{Q^2}{x}\right).
\label{(22)}
\ee
The functional behaviour of $\Lambda^2 (W^2) = \Lambda^2 \left( \frac{Q^2}{x}
\right)$ responsible for the $\vec r^2_\perp \rightarrow 0$ dependence
of the colour dipole cross section thus determines (or provides a model
for) the gluon density.

One may ask the additional question whether (\ref{(22)}) is consistent
with the result obtained by assuming the validity of the DGLAP evolution
\cite{11} in
application to our result for $F_2 = (Q^2 / 4 \pi^2 \alpha)\sigma_{\gamma^*
p}$. This is indeed the case. Assuming dominance of the gluon distribution at
low $x$ \cite{9}, we have with 4 quark flavours contributing,
\be
\frac{\partial F_2 \left( \frac{x}{2} , Q^2 \right)}{\partial \ln Q^2} =
\frac{10}{27\pi} x g (x) \alpha_s (Q^2) .
\label{(23)}
\ee
Using the $\eta \gg 1$ approximation in (\ref{(15)}) and noting that
(\ref{(3)}) yields
\be
\frac{\partial\Lambda^2 \left( \frac{Q^2}{x}\right)}{\partial \ln Q^2} =
C_2 \Lambda^2 \left( \frac{Q^2}{x} \right) ,
\label{(24)}
\ee
one immediately finds that (\ref{(23)})
yields a result for the gluon density that is identical to the one
in (\ref{(22)}).
In other words, the gluon density derived from the GVD/CDP
fulfills the evolution equation
(\ref{(23)}).
This result explicitly demonstrates the consistency of
the GVD/CDP with the DGLAP approach in the limit of $\eta \gg 1$, where
evolution is applicable.

In Fig.5, we show the gluon density based on (\ref{(22)})\footnote{
We note that (\ref{(20)}) as well as (\ref{(23)}), and accordingly also
(\ref{(22)}), are LO QCD relations. Consequently, also the extracted gluon
density in Fig.5 is a LO QCD result.}
using the evolution of $\alpha_s (Q^2)$ from the PDG \cite{10}.
For Fig.5, for $\sigma^{(\infty)}$
in
(\ref{(22)}), the value of $\sigma^{(\infty)} = 48~{\rm GeV}^{-2} = 18.7$ mb
was inserted, corresponding to the correct normalization of
$\sigma_{\gamma^* p}$ for the case of four active flavours.

Having established the connection between the colour-dipole cross section
of the GVD/CDP and the gluon density with evolution in $Q^2$, we are able
to estimate the domains in the $(Q^2 , x)$ and $(W^2 , x)$ planes for
which the evolution equation (\ref{(23)})
breaks down.
As noted, the transition from the region where the $\Lambda^2 \cdot
r^2_\perp \rightarrow 0$ approximation for the colour-dipole cross section
holds, to the region where its behaviour for $\Lambda^2 \cdot r^2_\perp
\rightarrow \infty$ becomes relevant, for the photoabsorption
cross section, $\sigma_{\gamma^* p} (\eta)$, corresponds to the
transition from large to small values of $\eta$. From Figs. 2 and 3,
the boundary of the two regions is approximately given by $\eta \cong 0.1$.
For $\eta \le 0.1$, the connection between the gluon density and the
colour-dipole cross section in (\ref{(20)}), (\ref{(22)}), and the evolution
equation (\ref{(23)}), breaks down. The curves in the $(Q^2 , x)$ and
$(W^2 , x)$ planes corresponding to $\eta = 0.1$ are shown in Fig.6.
From Fig.6, for e.g. $Q^2 = 0.5~{\rm GeV}^2$, the breakdown of DGLAP
evolution occurs for $x \le 10^{-5}$.

\bigskip

In conclusion, the GVD/CDP with scaling in $\eta$ provides a unique picture
for
the photabsorption cross section in the diffraction region of small values of
$x$, i.e. any $Q^2, x\rightarrow 0$.
The GVD/CDP covers the kinematic domain where the concept of a gluon density
evolving with $Q^2$ makes sense, as well as the kinematic domain, where
$\sigma_{\gamma^* p} \sim \ln (1 / \eta)$ and the limiting behaviour of
$\sigma_{\gamma^* p} / \sigma_{\gamma p} \rightarrow 1$ (``saturation'')
sets in. Scaling in $\eta$ allows for an important and fairly reliable
estimate of the boundary between these two kinematic domains.

\bigskip\noindent
{\bf Acknowledgement}\\
It is a pleasure to thank Wilfried Buchm\"uller, Leo Stodolsky
and G\"unter Wolf for useful
discussions. One of us (D.S.) thanks the theory group of the MPI f\"ur Physik
in M\"unchen for warm hospitality, where this work was finalized.

\newpage

\noindent

\medskip
\noindent
\noindent
\begin{figure}[ht]
\begin{minipage}[b]{.49\linewidth}
 \centering\epsfig{file=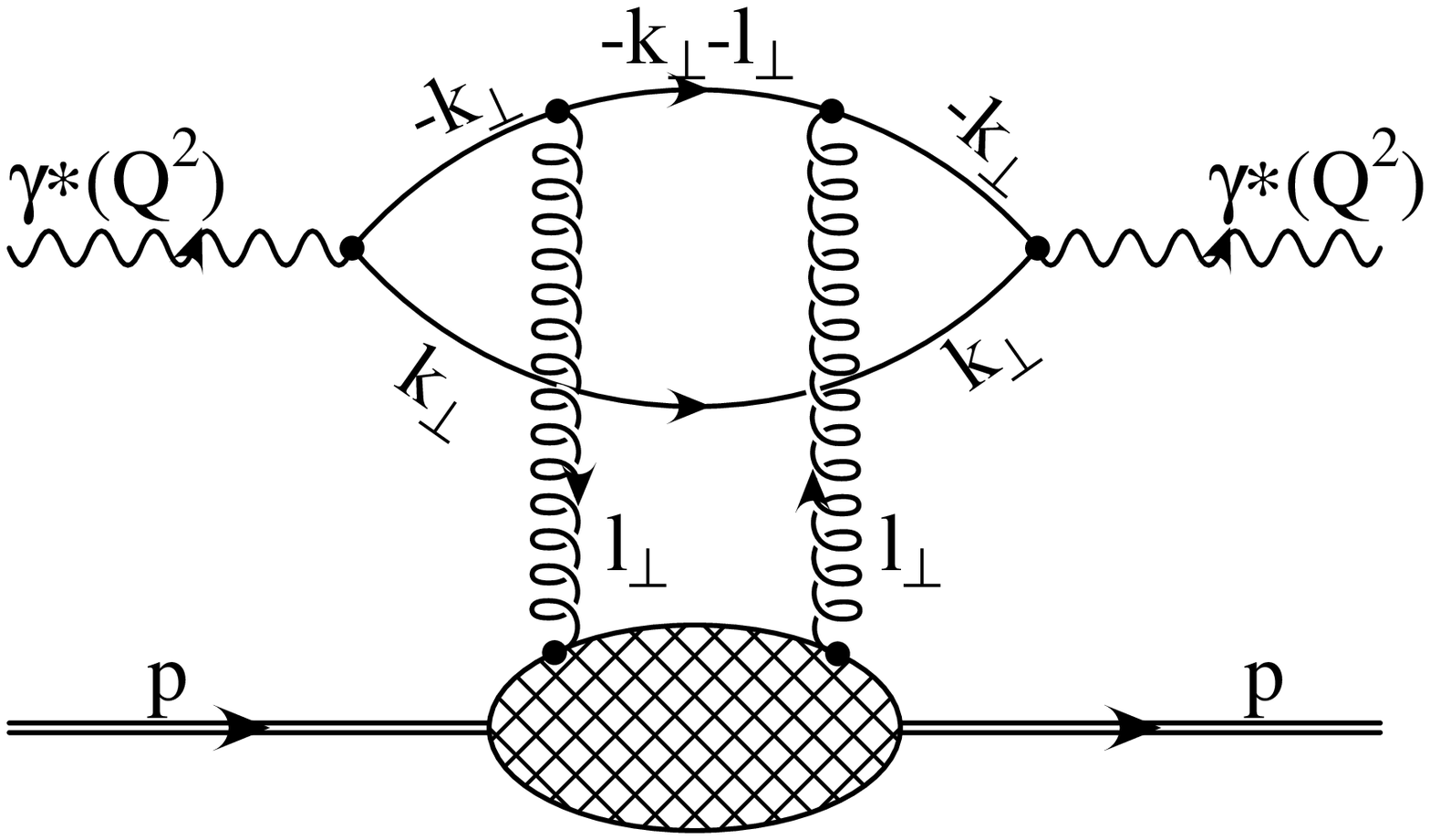,width=\linewidth}
\end{minipage}
\begin{minipage}[b]{.49\linewidth}
 \centering\epsfig{file=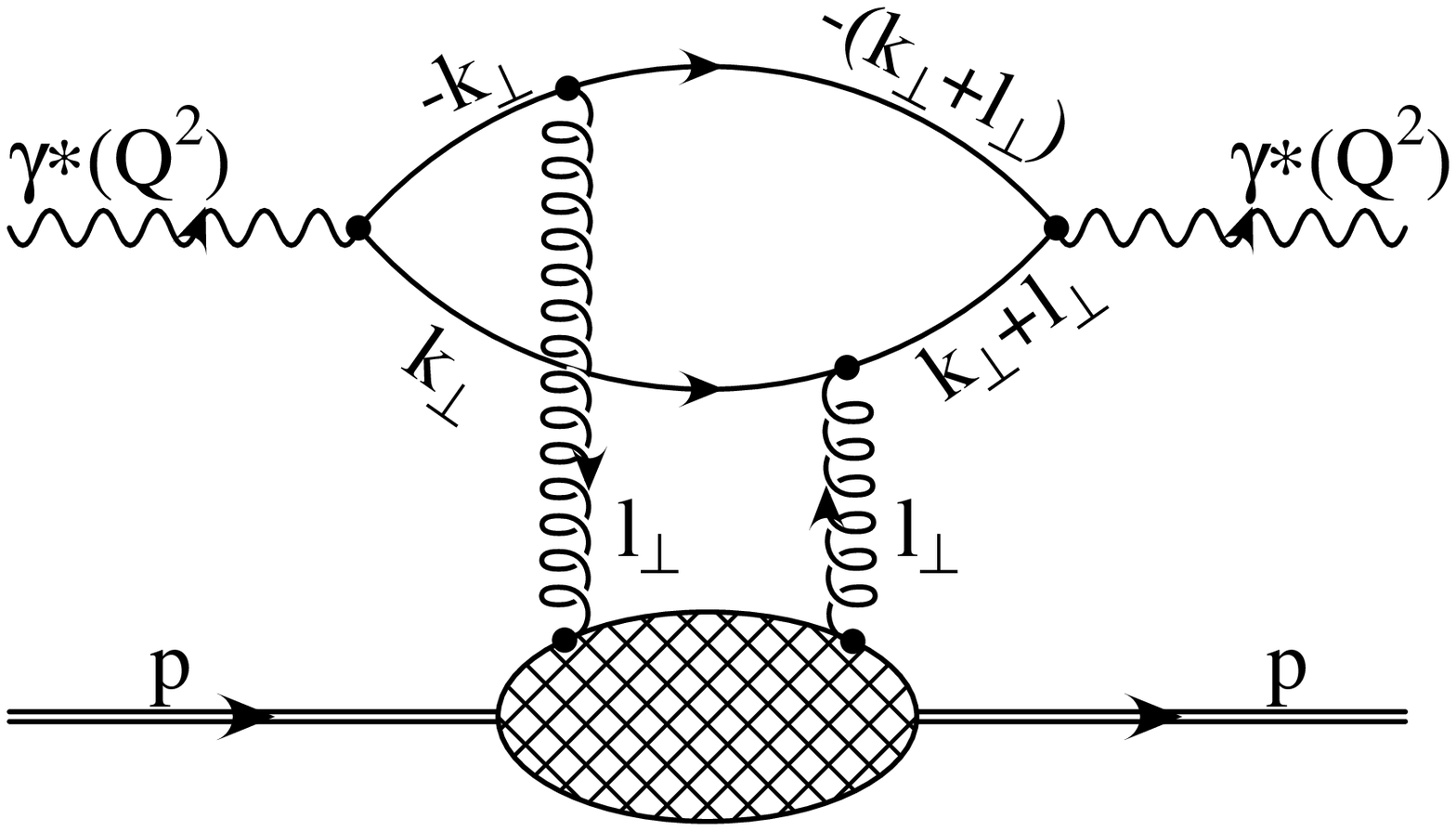,width=\linewidth}
\end{minipage}
\vspace{-1.0cm}
\caption{The two-gluon exchange.
The arrows relate to the transverse-momentum flow.
   \label{fig1}}
\end{figure}


\begin{figure}[ht]
\begin{center}
{\centerline{\epsfig{file=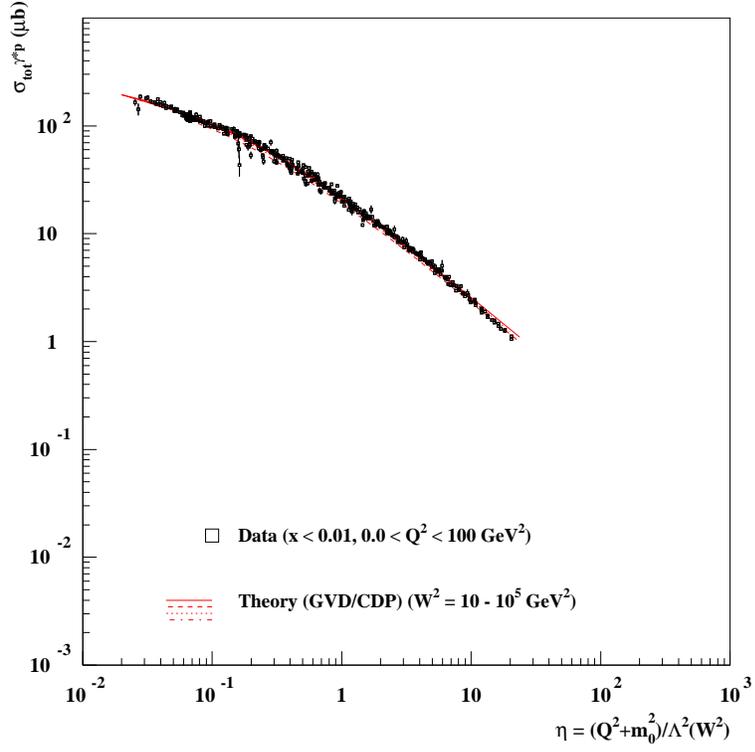,width=10.0cm}}}
\caption{
The virtual-photon-proton cross section, $\sigma_{\gamma^* p} (\eta (W^2,
Q^2))$ including $Q^2 = 0$ photoproduction as a function of $\eta =
(Q^2 + m^2_0) / \Lambda^2 (W^2)$.
\label{fig2}}
\end{center}
\end{figure}


\begin{figure}[ht]
\begin{center}
{\centerline{\epsfig{file=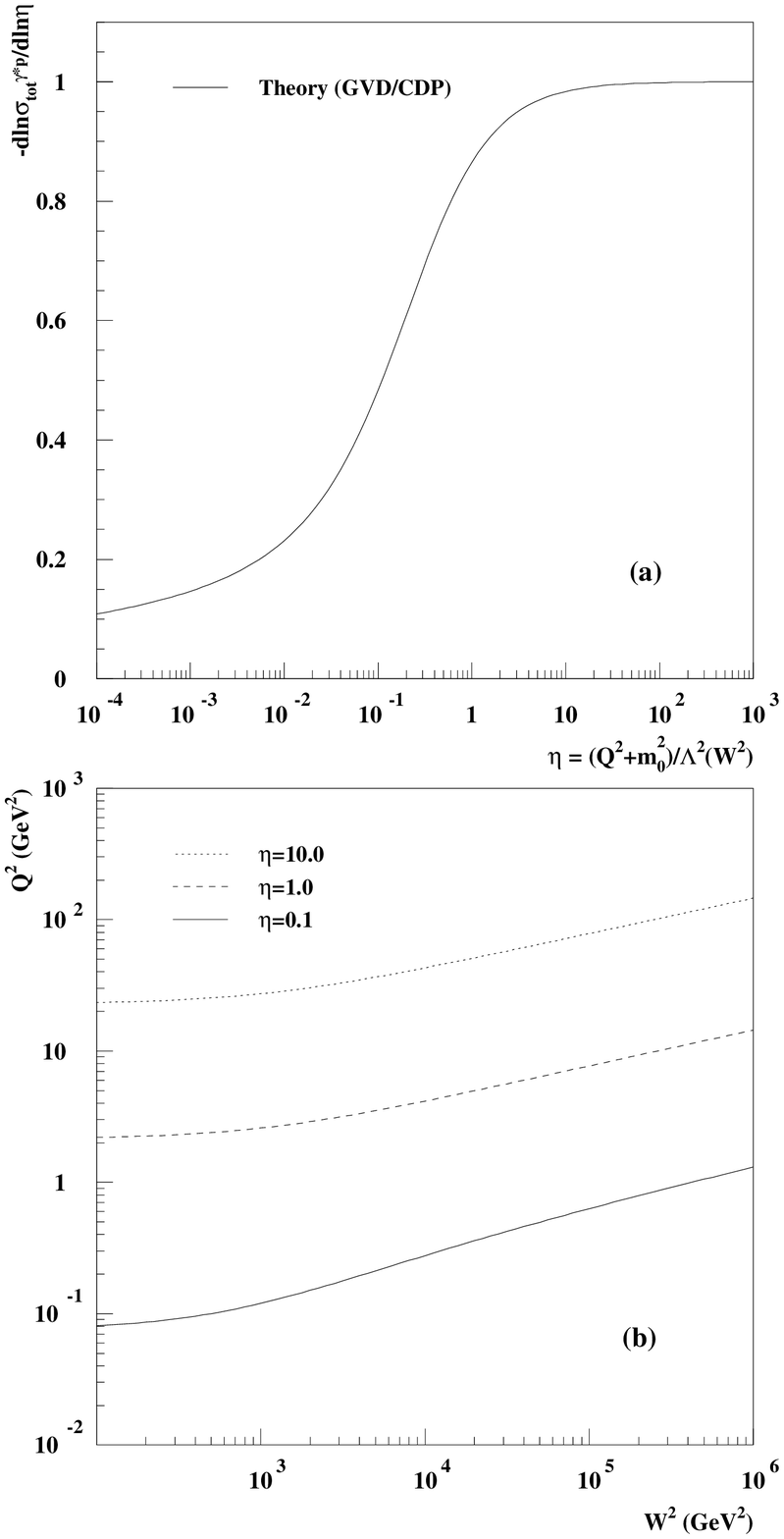,width=9.0cm}}}
\caption{
\newline
{\bf a)}
The logarithmic derivative of $\sigma_{\gamma^* p}$ as a function of
$\eta$.\newline
{\bf b)}
The lines of $\eta = $ const in the $(Q^2 , W^2)$ plane
\label{fig3}}
\end{center}
\end{figure}


\begin{figure}[ht]
\begin{center}
{\centerline{\epsfig{file=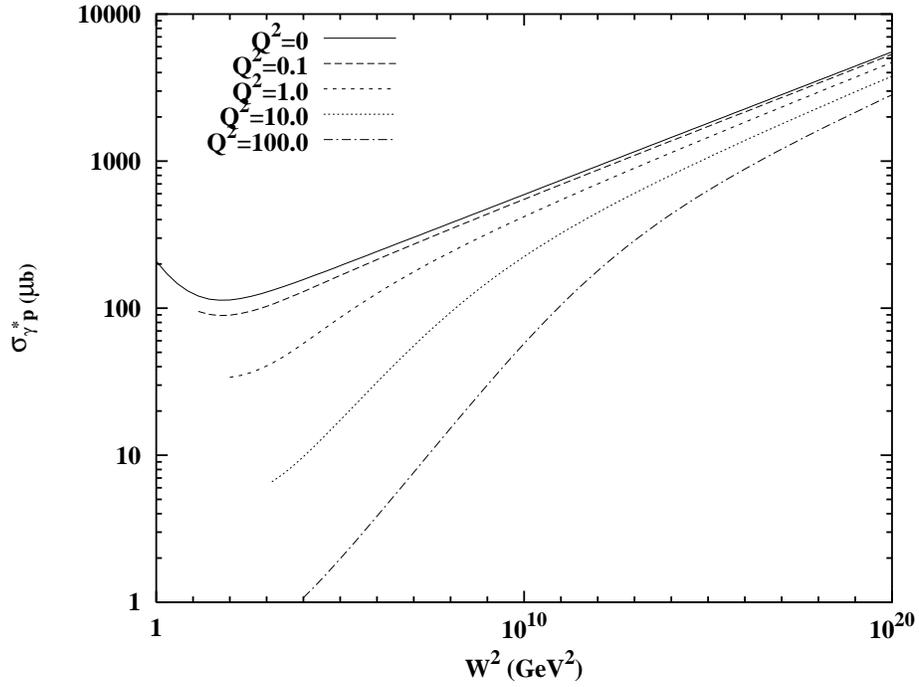,width=12.0cm}}}
\caption{
The virtual-photon-proton cross section, $\sigma_{\gamma^* p} (W^2 , Q^2)$,
including $Q^2 = 0$ photoproduction, as a function of $W^2$ for fixed
$Q^2$. The figure demonstrates the asymptotic behaviour,
$\sigma_{\gamma^* p} (W^2 , Q^2) / \sigma_{\gamma p} (W^2) \rightarrow 1$ for
$W^2 \rightarrow \infty$.
\label{fig4}}
\end{center}
\end{figure}


\begin{figure}[ht]
\begin{center}
{\centerline{\epsfig{file=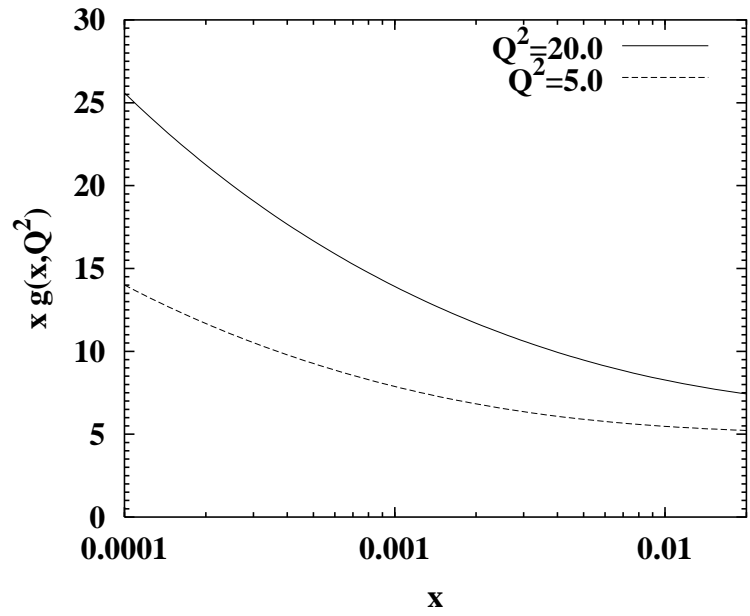,width=10.0cm}}}
\caption{
The gluon density corresponding to the colour-dipole cross section
of the GVD/CDP.
\label{fig5}}
\end{center}
\end{figure}

\newpage

\begin{figure}[ht]
 \centering\epsfig{file=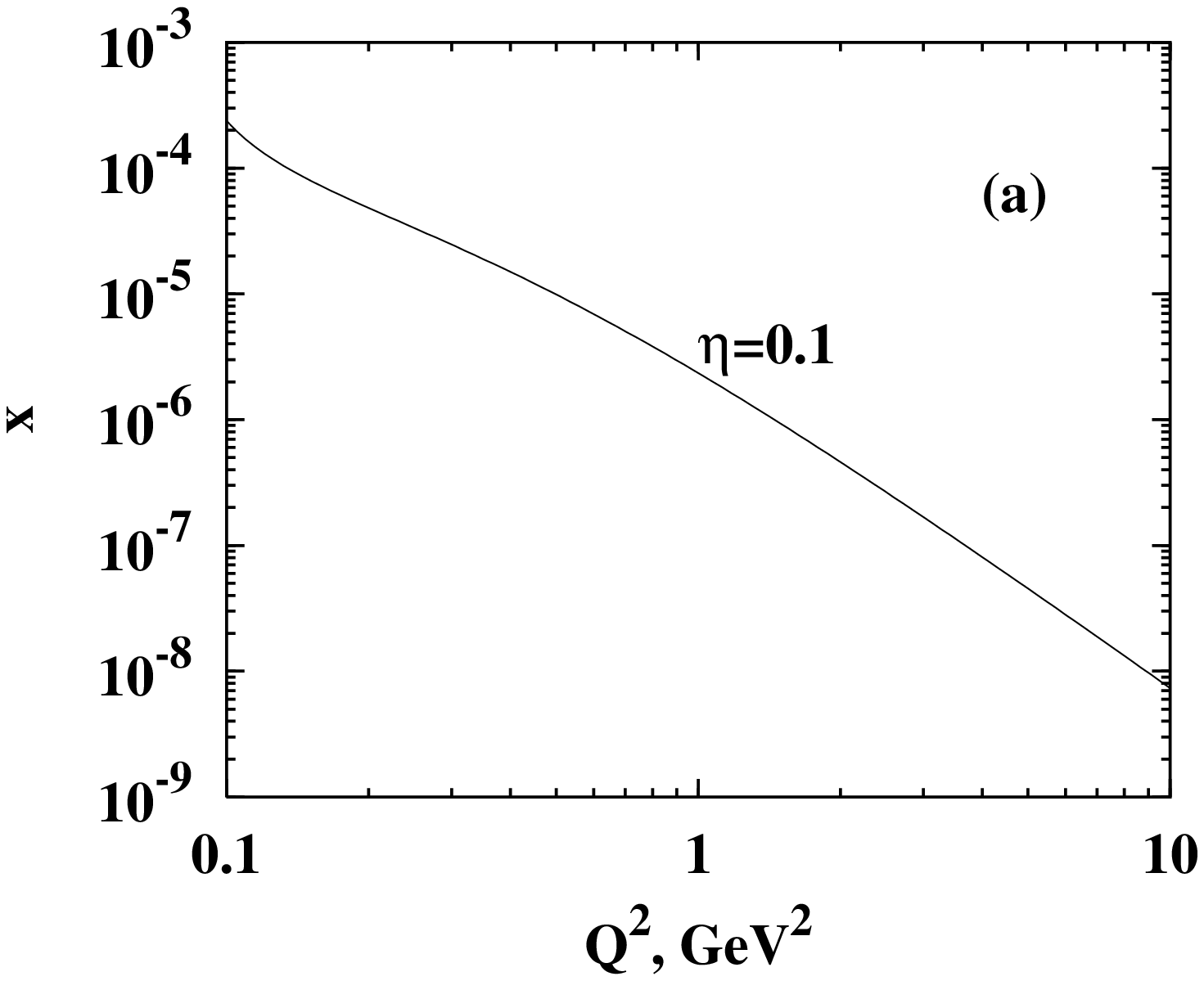,width=0.7\linewidth}
 \centering\epsfig{file=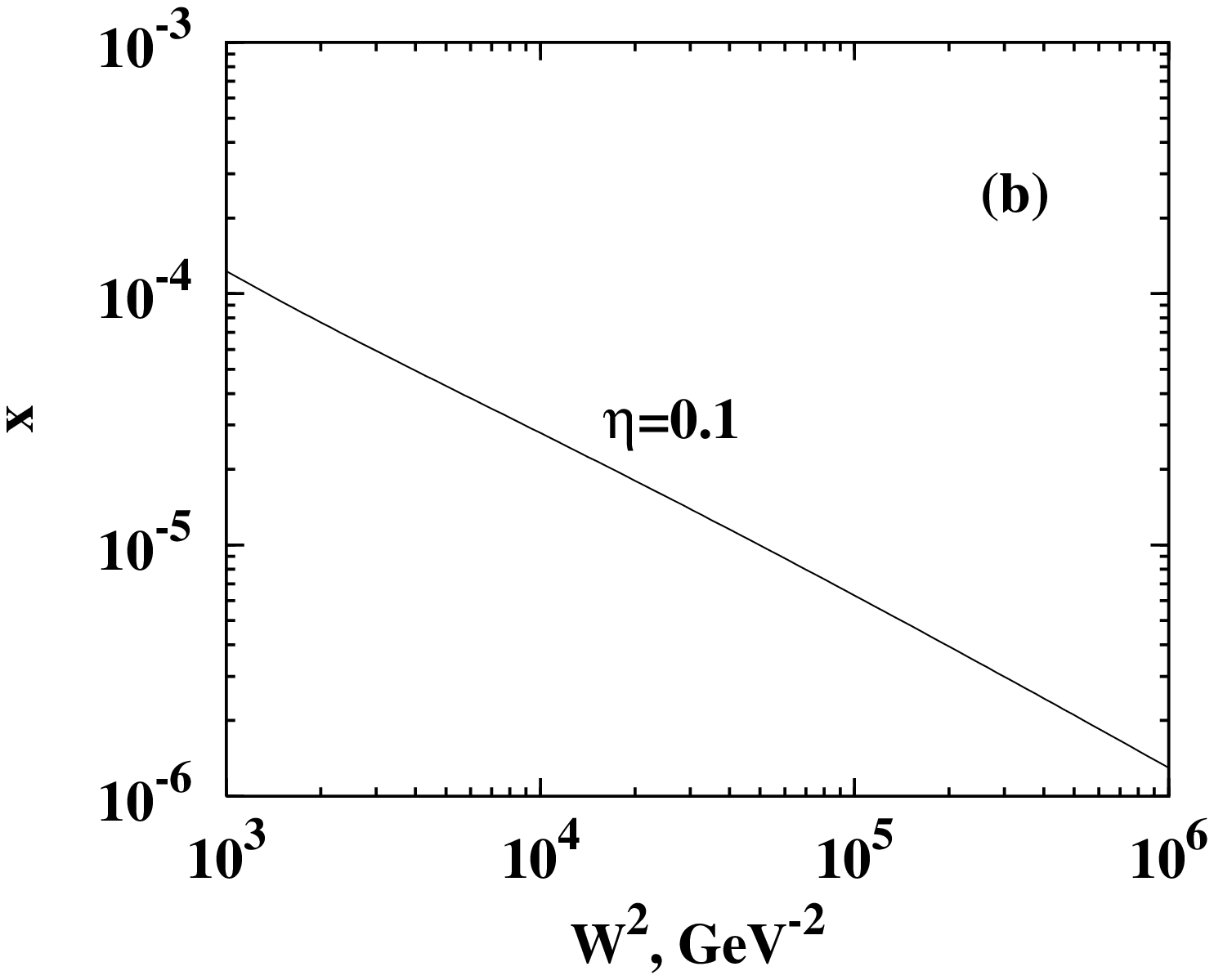,width=0.7\linewidth}
\caption{
The lines $\eta=0.1$ in the $(Q^2,x)$ plane (a) and in the $(W^2,x)$ plane
(b). The values of $x$ below the lines $\eta=0.1$ define the region, where
DGLAP evolution breaks down.
   \label{fig6}}
\end{figure}

\end{document}